\documentclass[twocolumn,aps,prl,showpacs,footinbib]{revtex4}%
\usepackage{graphicx}
\usepackage{amsmath}
\usepackage{amsfonts}
\usepackage{amssymb}%
\def\be{\begin{equation}}
\def\ee{\end{equation}}
\begin{document}
\title{Strongly enhanced light-matter interaction in a hybrid photonic-plasmonic resonator}
\author{Yun-Feng Xiao}
\altaffiliation{URL: www.phy.pku.edu.cn/$\sim$yfxiao/index.html}

\author{Yong-Chun Liu}
\email{ycliu@pku.edu.cn}
\author{Bei-Bei Li}
\author{You-Ling Chen}
\author{Yan Li}
\author{Qihuang Gong}
\email{qhgong@pku.edu.cn}
\affiliation{State Key Lab for Mesoscopic Physics, Department of Physics, Peking
University, P. R. China}
\date{\today}

\begin{abstract}
We propose a hybrid photonic-plasmonic resonant structure which consists of a metal nanoparticle (MNP) and a whispering
gallery mode (WGM) microcavity. It is found that the hybrid mode enables a strong
interaction between the light and matter, and the single-atom cooperativity is enhanced by more than two
orders of magnitude compared to that in a bare WGM microcavity. This remarkable improvement originates from
two aspects: (1) the MNP offers a highly enhanced local field in the vicinity
of an emitter, and (2), surprisingly, the high-\textit{Q} property of WGMs can
be maintained in the presence of the MNP. Thus the present system has great
advantages over a single microcavity or a single MNP, and holds great
potential in quantum optics, nonlinear optics and highly sensitive biosening.

\end{abstract}

\maketitle

Owing to the size mismatch between light and single emitters such as single
atoms, the interaction between them is very weak, so that it is of importance
to create a light-matter interface enabling strong interactions. One way to
bridge this mismatch is to employ the strong interaction within cavity quantum
electrodynamics (QED) \cite{CQED2002,CQED2005}. Cavity QED offers an almost
ideal platform for the study of physics at the interface of classical and
quantum mechanics, and provides a technology for various devices in the field
of quantum information \cite{QE,sqc,dqn}. Experiments on strong coupling
regime in cavity QED have made great advances over the past two decades
\cite{SCReview}. Among them, whispering gallery mode (WGM) microcavities
\cite{WGMQED} are promising because they possess ultrahigh quality ($Q$)
factor and allow for mass production on a chip. However, the relatively large
cavity mode volume makes it difficult to realize strong coupling. On the other
hand, due to the localized surface plasmon resonance (LSPR)
\cite{LSPRreview2004}, metal nanoparticles (MNPs) \cite{MNPreview2008} enable
subwavelength confinement of the optical field
\cite{Zhang2006,Nano2008,PRL2010,ACSNano2010,Waks2010}. Unfortunately, MNPs
suffer from serious absorption and scattering losses.

Against this backdrop, in this Letter, taking advantages from both
ultralow-loss WGMs and highly localized plasmon, we propose a WGM
microcavity-MNP resonant system. In this composite system, the high-$Q$ WGM
microcavity serves as a low-loss storage of the optical field, while the MNP
plays the role of an optical antenna which creates a hot spot and magnifies
the local optical field. Remarkably, the high-$Q$ property of WGMs can be
maintained in the presence of the MNP. As a result, the cooperativity
parameter (defined as $C=2G^{2}/\kappa\gamma_{s}$ \cite{Walls}, with $G$ being
the single photon coupling strength, $\gamma_{s}$ the spontaneous decay of the
emitter and $\kappa$ the decay of the cavity field) achieves a more than
$100$-fold increase compared with that of the WGM cavity alone. It should be
pointed out that, this composite cavity QED structure is significantly
different from previous designs where a silica disk or toroid was completely
covered with a metal layer, which led to strong degrading of the $Q$-factor
\cite{Min,Xiao}.

\begin{figure}[tb]
\begin{center}
\centerline{\includegraphics[width=\columnwidth]{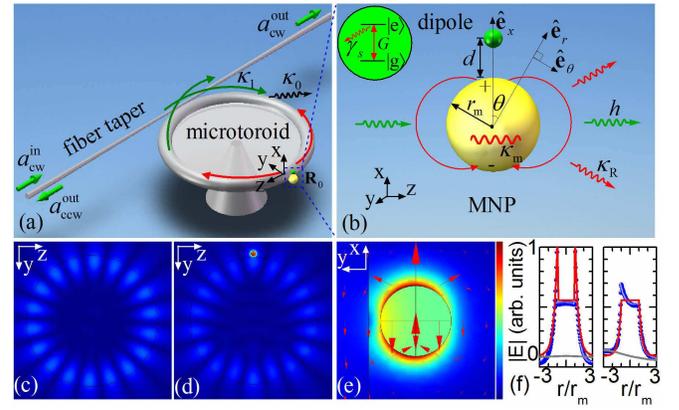}}
\end{center}
\caption{(Color online) (a) Sketch of the MNP-WGM composite system (not to
scale). (b) Zoom-in view of the MNP and the dipole emitter. The inset is the
energy diagram of the dipole. (c)-(e) Three-dimensional finite element
simulations of the electric field distribution. Top view of the electric field
profile without (c), and with (d) the MNP. (e) Zoom-in view of the field
distribution near the MNP (azimuthal cross-section). The red arrows show the
electric field directions. The vertical line on the left of the MNP is the
cavity boundary. The colors in (c)-(e) represent the norm of the electric
field ($\left\vert E\right\vert $). (f) Comparison of the electric field with
(blue dot) and without (light gray dot) MNP for $\theta=0$ (left) and $\theta
=\pi/2$ (right). The red solid curves are the theoretical result (see text). To save
simulation resources, here we use a relatively small microcavity (major radius
0.8 $\mathrm{\mu}$\textrm{m}) and a relatively large MNP ($r_{m}=30$
\textrm{nm}). }%
\label{fig1}%
\end{figure}

Figure \ref{fig1} illustrates a schematic of the system. A MNP is located onto
the surface of a microtoroidal cavity \cite{Vahala} which supports twin
counter-propagating WGMs with degenerate frequency ${\omega}_{\mathrm{c}}$
(neglecting the coupling between these two modes arising from surface
roughnesses induced scattering). A dipole emitter (ground state $\left\vert
g\right\rangle $, excited state $\left\vert e\right\rangle $, energy spacing
${\hbar}\omega_{\mathrm{e}}$ and dipole moment $\mu\mathbf{\hat{e}}_{x}$) is
placed in the vicinity of the MNP. Experimentally we can use atomic force
microscope (AFM) manipulation to controllably position the MNP and the dipole
emitter \cite{NL2011,NPhoton2008}. Alternatively, we can also use gold-coated
fiber taper tip to play the role of the MNP, in analogy to the method
utilizing fiber tip to simulate dielectric nanosphere \cite{Mazzei2007}. A
tapered fiber is used to couple light into and out of the microcavity. We
consider TE-polarized WGMs, with the dominant electric field component in the
$\mathbf{\hat{e}}_{x}$ direction. In the absence of the MNP, the
Jaynes-Cummings Hamiltonian reads $H_{\mathrm{c}}={\omega}_{\mathrm{c}}%
{\sum\nolimits_{n}}a_{n}^{\dag}a_{n}+\frac{1}{2}\omega_{\mathrm{e}}%
\,\sigma_{z}+G_{\mathrm{c}}{\sum\nolimits_{n}(}a_{n}^{\dag}\sigma_{-}%
+a_{n}\sigma_{+})$, where $\sigma_{-}{=}\sigma_{+}^{\dag}{=}\left\vert
g\right\rangle \left\langle e\right\vert $, ${\sigma_{z}=}\left\vert
e\right\rangle \left\langle e\right\vert -\left\vert g\right\rangle
\left\langle g\right\vert $, $a_{n}$ ($a_{n}^{\dagger}$) denotes annihilation
(creation) operator of the WGMs with $n=$ \textrm{CW (CCW)} corresponding to
the clockwise (counterclockwise) propagating mode. The single photon coupling
strength (vacuum Rabi frequency) is given by $G_{\mathrm{c}}=\mu
f_{\mathrm{c}}(\mathbf{R})({\omega}_{\mathrm{c}}/({2\hbar\varepsilon
_{\mathrm{0}}\varepsilon_{\mathrm{c}}V_{\mathrm{c}}))}^{1/2}$, where
${\varepsilon_{\mathrm{0}}}$ is the permittivity of vacuum, ${\varepsilon
_{\mathrm{c}}}$ is the relative permittivity of the microcavity,
${V_{\mathrm{c}}}$ and $f_{\mathrm{c}}(\mathbf{R})$ denote the mode volume and
normalized field distribution of the WGMs, respectively.

Now we focus on the case that a spherical MNP is placed in the vicinity of the
microcavity and the dipole emitter, with the position $\mathbf{R=R}%
_{\mathrm{0}}$. The radius of the MNP $r_{\mathrm{m}}$ is much smaller than
the light wavelength so that the interaction arising from the MNP is governed
by electrostatics rather than electrodynamics \cite{JPCB2003}, which is
confirmed by our numerical simulations (Fig. 1(c)-(f)). By solving Laplace's
equation, we obtain the positive frequency component of the total field inside
and outside the MNP: $\mathbf{E}_{\mathrm{c,m}}^{(+)}(r,\theta)=(1-\beta
)E_{\mathrm{c}}^{(+)}(\mathbf{R}_{\mathrm{0}})\mathbf{\hat{e}}_{x},(r\leq
r_{\mathrm{m}})$, $\mathbf{E}_{\mathrm{c,m}}^{(+)}(r,\theta)=E_{\mathrm{c}%
}^{(+)}(\mathbf{R}_{\mathrm{0}})\mathbf{\hat{e}}_{x}+\beta(r_{\mathrm{m}}%
^{3}/r^{3})E_{\mathrm{c}}^{(+)}(\mathbf{R}_{\mathrm{0}})(2\mathbf{\hat{e}}%
_{r}\cos\theta+\mathbf{\hat{e}}_{\theta}\sin\theta),(r>r_{\mathrm{m}})$. Here
the dipole emitter-induced field has been omitted in the weak excitation
limit; $r=\left\vert \mathbf{R-R}_{\mathrm{0}}\right\vert $ and $\theta$ form
a polar coordinates system; the complex coefficient $\beta$ is given by
$\beta=(\mathcal{\varepsilon}_{\mathrm{m}}-\mathcal{\varepsilon}_{\mathrm{b}%
})/(\mathcal{\varepsilon}_{\mathrm{m}}+2\mathcal{\varepsilon}_{\mathrm{b}})$,
where $\mathcal{\varepsilon}_{\mathrm{m}}$ (${\varepsilon_{\mathrm{b}}}$)
denotes the relative permittivity of the MNP (the surrounding medium). We can
use the Drude dispersion relation $\mathcal{\varepsilon}_{\mathrm{m}}%
(\omega)=1-\omega_{\mathrm{p}}^{2}/[\omega(\omega+i\gamma_{\mathrm{m}})]$,
where $\omega_{\mathrm{p}}$ is the bulk plasma frequency and $\gamma
_{\mathrm{m}}$\ is the damping rate which accounts for energy dissipation due
to ohmic losses.\ The LSPR \cite{LSPRreview2004} occurs with the dipolar
plasmonic resonance frequency $\omega_{\mathrm{sp}}$ satisfying
$\operatorname{Re}[\mathcal{\varepsilon}_{\mathrm{m}}(\omega_{\mathrm{sp}%
})]=-2\mathcal{\varepsilon}_{\mathrm{b}}$. In this case, $\left\vert
\beta\right\vert \gg1$, and the local field in the vicinity of the MNP is
enhanced significantly.

The response of the MNP results in the re-distribution of the cavity mode
field, forming a hybrid photonic-plasmonic mode. The modified maximum field
strength reads $\left\vert E_{\mathrm{c,m,\max}}\right\vert =f_{\mathrm{c}%
}(\mathbf{R}_{\mathrm{0}})\left\vert (1+2\beta)E_{\mathrm{c,\max}}\right\vert
$, and thereby the mode volume decreases to be $V_{\mathrm{c,m}}%
=\mathcal{\varepsilon}_{\mathrm{c}}V_{\mathrm{c}}/[\mathcal{\varepsilon
}_{\mathrm{b}}\left\vert 1+2\beta\right\vert ^{2}f_{\mathrm{c}}^{2}%
(\mathbf{R}_{\mathrm{0}})]$, which approximately scales as $\left\vert
\beta\right\vert ^{2}$ times smaller than bare cavity case. Thus the MNP
effectively plays the role of an optical antenna which (i) confines the photon
energy, (ii) reduces the mode volume, and (iii) magnifies the local field. For
the longitudinal case ($\theta=0$), the modified single photon coupling
strength $G_{\mathrm{c,m}}$ is calculated to be $\left\vert 1+2\beta
r_{\mathrm{m}}^{3}/r^{3}\right\vert $ times as large as $G_{\mathrm{c}}$.

In this composite cavity QED system, a key question is whether the MNP
significantly degrades the high-$Q$ property of WGMs. The MNP-induced decay
includes two contributions: scattering and absorption losses. For the
subwavelength MNP, the scattering interaction can be modeled in a dipole
approximation, where the electric field of the input wave induces a dipole
moment in it \cite{Mazzei2007}. On one hand, this scattering interaction mixes
the two counter-propagating modes, described by the Hamiltonian $H_{1}%
=-\frac{1}{2}\mathbf{p}_{\mathrm{m}}\cdot\mathbf{E}_{\mathrm{c,m}}%
(\mathbf{R}_{\mathrm{0}})$, where $\mathbf{p}_{\mathrm{m}}%
=\mathcal{\varepsilon}_{\mathrm{0}}\mathcal{\varepsilon}_{\mathrm{b}}\alpha
E_{\mathrm{c}}^{(+)}(\mathbf{R}_{\mathrm{0}})\mathbf{\hat{e}}_{x}+H.c.$ is the
polarization and $\alpha=4\pi r_{\mathrm{m}}^{3}\beta$ is the polarizability
of the spherical MNP. It can be simplified as $H_{1}={h}\sum
\nolimits_{n,n^{\prime}}a_{n}^{\dag}a_{n^{\prime}}$, where $n$, $n^{\prime}=$
\textrm{CW}, \textrm{CCW}, and $h=2\pi r_{\mathrm{m}}^{3}\mathcal{\varepsilon
}_{\mathrm{b}}{\omega}_{\mathrm{c}}\left\vert \beta\right\vert ^{2}%
f_{\mathrm{c}}^{2}(\mathbf{R}_{\mathrm{0}})/(\mathcal{\varepsilon}%
_{\mathrm{c}}{V_{\mathrm{c}}})$ is the coupling strength. On the other hand,
the scattering results in the decay from WGMs to reservoir modes, described by
the Hamiltonian $H_{2}=-\frac{1}{2}\sum\nolimits_{j}(\mathbf{p}_{\mathrm{m}%
}\cdot\mathbf{E}_{j}(\mathbf{R}_{\mathrm{0}})+\mathbf{p}_{j}\cdot
\mathbf{E}_{\mathrm{c,m}}(\mathbf{R}_{\mathrm{0}}))$, where $\mathbf{E}%
_{j}(\mathbf{R}_{\mathrm{0}})$ is the electric field for the $j$th reservoir
mode and $\mathbf{p}_{j}$ is reservoir-induced polarization of the MNP. Taking
the Weisskopf-Wigner semi-QED treatment \cite{Mazzei2007}, the scattering
process results in radiation energy decay of the WGMs, which takes an
analogous form of spontaneous emission. After tracing out the freedom of the
reservoir modes, we obtain the effective Hamiltonian $H_{\mathrm{2,eff}%
}=-i\kappa_{\mathrm{R}}\sum\nolimits_{n,n^{\prime}}a_{n}^{\dag}a_{n\prime}$
with the damping rate $\kappa_{\mathrm{R}}=\mathcal{\varepsilon}_{\mathrm{b}%
}^{5/2}(4\pi r_{\mathrm{m}}^{3})^{2}\left\vert \beta\right\vert ^{4}{\omega
}_{\mathrm{c}}^{4}f_{\mathrm{c}}^{2}(\mathbf{R}_{\mathrm{0}})/(6\pi
c^{3}\mathcal{\varepsilon}_{\mathrm{c}}{V_{\mathrm{c}}})$, where $c$ is the
speed of light in vacuum.

The absorption of the MNP results in ohmic losses, characterized by the
imaginary part of the metal's permittivity. Thus this energy dissipation can
be described by the non-Hermitian Hamiltonian $H_{\mathrm{3,eff}}=\frac{i}%
{2}\mathcal{\varepsilon}_{\mathrm{0}}\int\nolimits_{\left\vert \mathbf{R-R}%
_{\mathrm{0}}\right\vert \leq r_{\mathrm{m}}}\operatorname{Im}\left[
d(\omega\mathcal{\varepsilon}_{\mathrm{m}}(\omega))/\omega\right]  \left\vert
_{\omega=\omega_{\mathrm{c}}}\right.  E_{\mathrm{c,m}}(\mathbf{R}%
)^{2}d\mathbf{R}^{3}$, re-written as $H_{\mathrm{3,eff}}=-i\kappa_{\mathrm{m}%
}\sum\nolimits_{n,n^{\prime}}a_{n}^{\dag}a_{n\prime}$, with the decay rate
$\kappa_{\mathrm{m}}=4\pi r_{\mathrm{m}}^{3}\left\vert 1-\beta\right\vert
^{2}\omega_{\mathrm{p}}^{2}\gamma_{\mathrm{m}}f_{\mathrm{c}}^{2}%
(\mathbf{R}_{\mathrm{0}})/(3{\varepsilon_{\mathrm{c}}\omega_{\mathrm{c}}%
^{2}V_{\mathrm{c}})}$. In current case, the energy fraction of the hybrid mode
within the MNP is very small ($<0.01\%$), resulting a minor $\kappa
_{\mathrm{m}}$.

\begin{figure}[tb]
\begin{center}
\includegraphics[width=7.5cm]{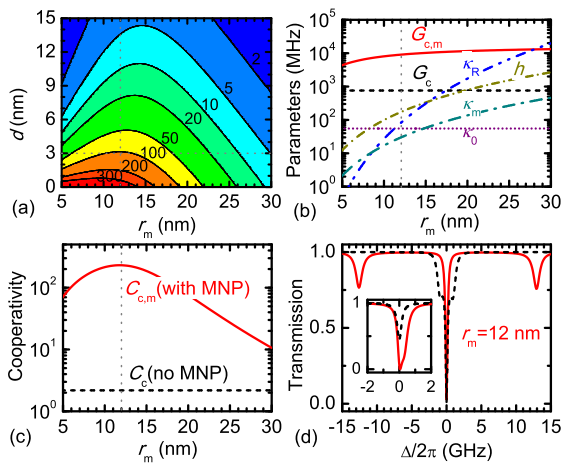}
\end{center}
\caption{(Color online) (a) Contour plot of the cooperativity enhancement
($C_{\mathrm{c,m}}/C_{\mathrm{c}}$) vs. $r_{\mathrm{m}}$ and $d$. (b)
Parameters \{$G_{\mathrm{c}}$, $G_{\mathrm{c,m}}$, $h$, $\kappa_{0}$,
$\kappa_{\mathrm{R}}$, $\kappa_{\mathrm{m}}$\}$/2\pi$ vs. $r_{\mathrm{m}}$.
(c) Cooperativity with the presence of the MNP ($C_{\mathrm{c,m}}$) and with
its absence ($C_{\mathrm{c}}$). The horizontal line in (a) corresponds to
$d=3$ \textrm{nm}. The vertical lines in (a)-(c) corresponds to $r_{\mathrm{m}%
}=12$ \textrm{nm}. (d) Transmission spectra for $r_{\mathrm{m}}=12$ \textrm{nm
}and\textrm{ }$d=3$ \textrm{nm }in the presence of the dipole, with (solid
line) and without the MNP (dashed line). \{$G_{\mathrm{c}}$, $G_{\mathrm{c,m}%
}$, $h$, $\kappa_{0}$, $\kappa_{\mathrm{R}}$, $\kappa_{\mathrm{m}}$\}$/2\pi
=$\{$760$, $9000$, $170$, $55$, $80$, $30$\} \textrm{MHz}. The input-cavity
detuning $\Delta=\omega-\omega_{\mathrm{c}}$. The inset is for no dipole case.
For (b)-(d), $d=3$ \textrm{nm}; For (a)-(d), the cavity-MNP detuning
$\Delta_{\mathrm{sp}}=0$.}%
\label{fig2}%
\end{figure}

Now we arrive at the overall Hamiltonian of the composite system
$H_{\mathrm{c,m}}=H_{\mathrm{c}}+H_{1}+H_{\mathrm{2,eff}}+H_{\mathrm{3,eff}}$.
Next we consider specific examples. Different from low-$Q$ photonic crystal
nanocavity \cite{Barth}, here we focus on the enhanced light-matter
interaction in a silica microtoroidal cavity \cite{Vahala} which possesses a
relatively high intrinsic quality factor ${Q}_{0}${\ }${=10}^{7}$. We use a
typical toroidal microcavity with the major and minor radii being $30$
$\mathrm{\mu m}$, $3$ $\mathrm{\mu m}$, respectively. For such a cavity in
air, ${\varepsilon_{\mathrm{c}}=1.45}^{2}$, ${\varepsilon_{\mathrm{b}}=1}$, we
obtain $V_{\mathrm{c}}\sim200$ $\mathrm{\mu m}^{\mathrm{3}}$ (corresponding to
$4000(\lambda/n)^{3}$) and $f_{\mathrm{c}}(\mathbf{R}_{\mathrm{0}})\sim0.3$
using finite element simulations. We set the cavity-taper coupling strength{
}$\kappa_{1}=5\kappa_{0}$, where ${\kappa_{0}=\omega_{\mathrm{c}}/Q}_{0}$
denotes the intrinsic damping of the WGMs{. We use large }${\kappa_{1}}$ here
so as to realize near critical coupling in the presence of the MNP which
brings about additional decays.{ }For a gold MNP, the permittivity can be
extracted from the experimental data \cite{PRB1972}, with $\omega_{\mathrm{p}%
}\sim6\times10^{15}$ \textrm{Hz} (corresponding to the LSPR wavelength of
$540$ \textrm{nm}) and $\gamma_{\mathrm{m}}\sim3\times10^{14\text{ }}%
$\textrm{Hz}. For the dipole emitter, we can use chemically synthesized
cadmium selenide (CdSe) quantum dot, with the dipole moment $\mu
=2.4\times10^{-28}$ \textrm{C}$\mathrm{\cdot}$\textrm{m }\cite{ACSNano2010},
the spontaneous emission rate $\gamma_{\mathrm{s}}=$ $\mathcal{\varepsilon
}_{\mathrm{b}}^{1/2}\mu^{2}\omega_{\mathrm{e}}^{3}/(3\pi\epsilon_{\mathrm{0}%
}\hbar c^{3})=2\pi\times1.6$ \textrm{GHz} and the emitter-cavity detuning
$\Delta_{\mathrm{ec}}\equiv\omega_{\mathrm{e}}-{\omega_{\mathrm{c}}=0}$. To
focus on the physics, the size of the quantum dot is assumed to be small so
that its mesoscopic effects \cite{NPhys2010} can be neglected. Note that for a
more general emitter like an atom or a dye molecule, the size can be
considered nil. In Fig. \ref{fig2}(a) we plot the cooperativity enhancement as
a function of $r_{\mathrm{m}}$ and $d$ (the distance between the dipole
emitter and the surface of the MNP). It is shown that the enhancement can
exceed more than two orders of magnitude, and there is enough parameter space
for remarkable enhancement. As an example, we set $d=3$ \textrm{nm},
corresponding to the horizontal line in Fig. \ref{fig2}(a). At this distance,
the charge carrier tunneling can be safely avoided \cite{NL2005,PRL2006}. With
a MNP of specific geometry \cite{OL2007}, the dipole's emission to dark
multipolar plasmonic resonances, which can be regarded as an addition
spontaneous decay of the dipole emitter and largely related to the concept of
the \textquotedblleft quenching\textquotedblright\ in the fluorescence
measurement, will be significantly reduced to a minor efficiency of $1\%-3\%$.
The metal absorption in the hybrid mode, also related to \textquotedblleft
quenching\textquotedblright, is extremely weak due to the ultra-small energy
fraction ($<0.01\%$) within the MNP, which differs from the pure plasmonic
resonant system. When LSPR occurs ($\Delta_{\mathrm{sp}}\equiv\omega
_{\mathrm{c}}-\omega_{\mathrm{sp}}$), we obtain $\left\vert \beta\right\vert
=11.5$. Figure \ref{fig2}(b) plots the characteristic parameters depending on
the radius of the MNP. It explicitly shows that the dipole-field single photon
coupling strength with the help of the MNP ($G_{\mathrm{c,m}}$, solid line)
can be more than $10$ times larger than that without the MNP ($G_{\mathrm{c}}%
$, dashed line), with achievable $G_{\mathrm{c,m}}/2\pi\simeq13$ \textrm{GHz}.
When $r_{\mathrm{m}}\ $is small, the extra decays ($\kappa_{\mathrm{R}}%
,\kappa_{\mathrm{m}}$) are smaller than the intrinsic cavity decay $\kappa
_{0}$. Therefore, for a small MNP, the enhancement of the local field can be
quite strong, while the high-$Q$ property of the system is surprisingly
maintained. In Fig. \ref{fig2}(c), we plot the cooperativity parameter as a
function of $r_{\mathrm{m}}$ both in the presence of the MNP and in its
absence. We can see that for $r_{\mathrm{m}}\simeq12$ \textrm{nm}, the
cooperativity is enhanced more than $100$-folds, with the corresponding
$G_{\mathrm{c,m}}/2\pi=9$ \textrm{GHz}.

To probe the enhanced light-matter interaction, we use weak monochromatic
clockwise propagating light input $a_{\mathrm{cw}}^{\mathrm{in}}$. By
utilizing the Heisenberg-Langevin equations and the input-output theory
\cite{Gardiner,Liu2011}, we obtain the cavity transmission spectra for
$r_{\mathrm{m}}=$ $12$ \textrm{nm} in Fig. \ref{fig2}(d). When the MNP is
absent, the vacuum Rabi splitting cannot be resolved. On the contrary, with
the help of the MNP, a mode splitting of $2\sqrt{2}G_{\mathrm{c,m}}$ can be
observed. This splitting is completely the cavity QED effect and the
scattering of the MNP has no contribution to it. To demonstrate this, in the
inset of Fig. \ref{fig2}(d), we plot the transmission spectra for the no
dipole emitter case. As expected, no splitting is observed, where the
asymmetry spectrum in the presence of the MNP is due to the scattering-induced
mixing of the WGMs.

\begin{figure}[ptb]
\begin{center}
\includegraphics[width=7.5cm]{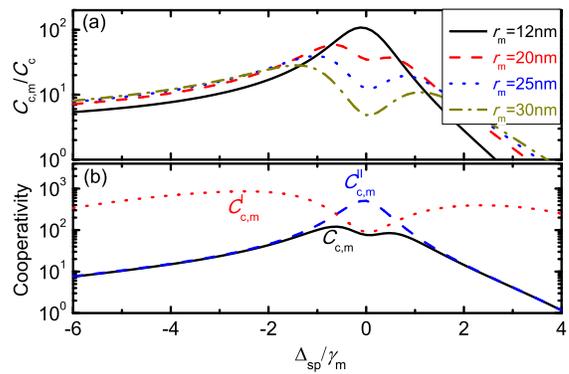}
\end{center}
\caption{(Color online) (a) Cooperativity enhancement ($C_{\mathrm{c,m}%
}/C_{\mathrm{c}}$) vs. cavity-MNP detuning $\Delta_{\mathrm{sp}}%
/\gamma_{\mathrm{m}}$ for several typical $r_{\mathrm{m}}$. (b) The
cooperativity parameters $C_{\mathrm{c,m}}$, $C_{\mathrm{c,m}}%
^{\text{\textrm{I}}}$ (near-resonance, $|\Delta_{\mathrm{sp}}|\ll\gamma_{m}$)
and $C_{\mathrm{c,m}}^{\text{\textrm{II}}}$ (off-resonance, $|\Delta
_{\mathrm{sp}}|\gg\gamma_{m}$) for $r_{\mathrm{m}}=20$ \textrm{nm.}}%
\label{fig3}%
\end{figure}

One distinct property of LSPR is that the plasmon excitation covers a broad
bandwidth. In fact, the optimal condition for the improvement of strong
coupling does not occur at the exact cavity-MNP resonance for large-size MNP
where the MNP-induced decay exceeds the original decay. This is verified in
Fig. \ref{fig3}(a), which displays the cooperativity parameters for cavity-MNP
detuning ($\Delta_{\mathrm{sp}}\equiv\omega_{\mathrm{c}}-\omega_{\mathrm{sp}}%
$) varying from $-6\gamma_{\mathrm{m}}$ to $4\gamma_{\mathrm{m}}$. It is
interesting that $C_{\mathrm{c,m}}/C_{\mathrm{c}}$ arrives at a local minimum
at $\Delta_{\mathrm{sp}}=0$ for a large $r_{\mathrm{m}}$. The underlying
physics is that the MNP-induced decay contributes more than the field
enhancement. To explain this more clearly, we plot the cooperativity for
$r_{\mathrm{m}}=20$ \textrm{nm }in Fig. \ref{fig3}(b). In the near-resonance
region, the loss induced by the MNP is dominant, and thereby $C_{\mathrm{c,m}%
}\simeq$ $C_{\mathrm{c,m}}^{\mathrm{I}}=2G_{\mathrm{c,m}}^{2}/[\gamma
_{\mathrm{s}}(\kappa_{\mathrm{R}}+\kappa_{\mathrm{m}})]$; while in the
off-resonance region, these decays decrease to a low level, yielding
$C_{\mathrm{c,m}}\simeq$ $C_{\mathrm{c,m}}^{\mathrm{II}}=2G_{\mathrm{c,m}}%
^{2}/[\gamma_{\mathrm{s}}(\kappa_{0}+\kappa_{1})]$. It is shown that, for red
cavity-MNP detuning case, the cooperativity can maintain a high value over a
broad range. Note that $\Delta_{\mathrm{sp}}=0,-6\gamma_{\mathrm{m}}$
corresponds to the wavelength $540\ \mathrm{nm}$, $1130$ $\mathrm{nm}$,
respectively. Moreover, even for $r_{\mathrm{m}}=30$ $\mathrm{nm}$, the
cooperativity can still be enhanced by $30$-folds for a suitable detuning.

\begin{figure}[tb]
\begin{center}
\centerline{\includegraphics[width=7cm]{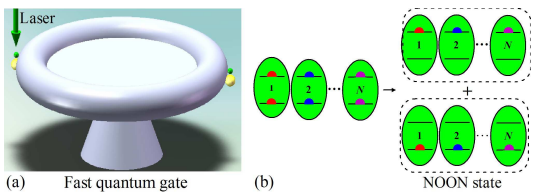}}
\end{center}
\caption{(Color online) Illustration of the fast quantum gate operation (a)
and NOON state generation (b).}%
\label{fig4}%
\end{figure}

Enhancing the strong coupling in cavity QED is of crucial importance in
high-speed operation of quantum gate and fast generation of entangled state,
as illustrated in Fig. \ref{fig4}(a). These can be obtained among two or more
identical MNP-atom molecules localized near the microcavity. The atoms
interact with a shared cavity mode and a classical laser field is applied to
control the evolution among them. Typically, to suppress the atomic
spontaneous emission, the system usually works in the large atom-field
detuning limit ($\Delta_{\mathrm{ec}}\gg G$) \cite{QE,Largedetuning}. For
conventional cavity QED setup without MNPs, this large detuning limit is
theoretically feasible, but may not be applicable because the effective
interacting strength ($\propto G^{2}/\Delta_{\mathrm{ec}}$) decreases
significantly and the Rabi oscillation becomes slow. On the contrary, with the
help of MNPs, the effective interaction is significantly enhanced. As an
example, the effective interacting strength can be improved $100$-folds with
the help of MNPs. This means that the quantum gate operation and the
entanglement generation can finish in a much shorter time, and the system
allows for much more operations within the coherence time.

Other examples include the fast generation of spin squeezed state
\cite{KU1993} and $N$ dipole emitters maximally entangled state (also known as
NOON state) \cite{NOON2010}. To this end, $N$ identical MNP-atom molecules are
positioned around the cavity. In the large detuning case, the effective
Hamiltonian is given by $H_{\mathrm{eff}}=\chi(J_{z}-J_{z}^{2})$, where
$\chi=2G_{\mathrm{c,m}}^{2}/(\Delta_{\mathrm{ec}}-Nh)$ and $J_{z}=\frac{1}%
{2}\sum_{j=1}^{N}{\sigma_{z}^{(j)}}$. This Hamiltonian is the generalized
one-axis-twisting type \cite{Jin2009}, which implies that it is capable of
performing spin squeezing. To create the NOON state, dipole emitters are
initially prepared in the superposition state $2^{-N/2}(\left\vert
g\right\rangle +\left\vert e\right\rangle )^{\otimes N}$ using a fast $\pi/2$
pulse. Next the system undergoes free evolution for $\chi t=\pi/2$. Then a
second $\pi/2$ pulse is applied to finally create the NOON state $(\left\vert
g\right\rangle ^{\otimes N}+\left\vert e\right\rangle ^{\otimes N})/\sqrt{2}$,
as described in Fig. \ref{fig4} (b). Previously this scheme is subjected to
the weak nonlinearity and\ experimentally it is a challenge to increase the
nonlinearity so as to overcome the decoherence. In our proposed composite
system the nonlinear coefficient $\chi$ can be large enough so that the
squeezing and the NOON state can be obtained in a short time, which is very
important in view of decoherence. Moreover, the additional dissipation induced
by the MNP has little effect on the system because here the dipole emitters
interact dispersively with the cavity modes.

In summary, we have studied a composite MNP-WGM coupling system. With the
help of the MNP, the single photon coupling strength between a dipole emitter
and the cavity system can be substantially enhanced, while the high-$Q$
property is surprisingly preserved. The single-atom cooperativity obtains a more
than two orders of magnitude increase compared with the conventional WGM
cavity QED system. Also, the MNP-induced coupling enhancement covers a broad
band up to several hundred nanometers so that the microcavity and the MNP do
not have to be on resonance. The system is applicable ranging from quantum
optics to quantum information science. It also offers new opportunities to
exploit both theoretical and experimental physics in stronger light-matter
interaction regime, such as nonlinear optics (e.g., enhanced Raman and
Rayleigh scattering) and highly sensitive biosening (e.g., single nanoparticle
sizing with an improved detection limit).

\begin{acknowledgments}
Y.F.X and Y.C.L. contribute equally. Y.F.X thanks Edo Waks for stimulating discussions. This work was supported by
NSFC (Nos. 11004003, 11023003, and 11121091), 973 program (No. 2007CB307001),
and RFDPH (No. 20090001120004).
\end{acknowledgments}

\end{document}